\documentclass[a4paper]{PoS}
\usepackage{microtype}

\title{Sensitivity of DUNE to long-baseline neutrino oscillation physics}

\ShortTitle{Sensitivity of DUNE to LBL oscillation physics}

\author{
   \speaker{Justo Mart\'in-Albo}\footnote{Now at Harvard University (Cambridge, MA, United States).}~ for the DUNE Collaboration\\
   Department of Physics, University of Oxford\\
   Denys Wilkinson Building, Keble Road, Oxford OX1 3RH, United Kingdom\\
   E-mail: \email{justo.martin-albo@physics.ox.ac.uk}}

\abstract{DUNE is an international project, currently in its design phase, for neutrino physics and proton-decay searches. It will consist of two detectors exposed to a megawatt-scale muon neutrino beam that will be produced at Fermilab (Illinois, USA). One detector will record particle interactions near the source of the beam, while a second, much larger, detector comprising four 10-kilotonne liquid argon TPCs will be installed at a depth of 1.5~km at SURF (South Dakota, USA), about 1300 kilometres away of the neutrino source. Among the primary scientific goals of DUNE is the precision measurement of the parameters that govern neutrino mixing, including those still unknown: the octant in which the $\theta_{23}$ mixing angle lies, the neutrino mass ordering and the value of the CP violation phase. This paper discusses the sensitivity of the DUNE experiment to these parameters.}

\FullConference{EPS-HEP 2017, European Physical Society conference on High Energy Physics\\ 5-12 July 2017\\ Venice, Italy}

\begin{document}

\section{Introduction}
%%%
Over the last two decades, oscillation experiments have established that neutrinos are massive particles and that the three neutrino states participating in the weak interaction ($\nu_\ell$, with $\ell = e, \mu, \tau$) are linear combinations of three mass states ($\nu_i$, with $i=1,2,3$) that control free-particle evolution:
%%%
\begin{equation}
\nu_\ell = \sum_i\ U^*_{\ell i}\ \nu_i\,.
\end{equation}
%%%
Here, $U$ is a $3\times3$ unitary matrix that is usually parametrized in terms of 3 Euler angles ($\theta_{12}$, $\theta_{13}$, $\theta_{23}$) and up to 3 phases ($\delta$, $\alpha_1$, $\alpha_2$), depending on the Dirac or Majorana nature of neutrino masses. The three mixing angles have been measured to varying degrees of accuracy, but the value of the Dirac CP-violation phase, $\delta$, also potentially measurable in neutrino oscillation experiments, is currently unknown. 

Oscillation experiments are not sensitive to the absolute neutrino mass scale; they can only measure the differences between the squared neutrino masses, $\Delta m^2_{ij} \equiv m^2_i - m^2_j$. Experiments using solar and reactor neutrinos have measured one mass difference, the so-called solar mass splitting: $\Delta m_{\rm sol}^2\simeq7.5\times10^{-5}$~eV$^2$. Atmospheric and accelerator-based oscillation experiments have measured another mass difference, the atmospheric mass splitting: $\Delta m_{\rm atm}^2\simeq2.5\times10^{-3}$~eV$^2 \gg \Delta m_{\rm sol}^2$. These results do not let us differentiate between two possibilities for the ordering of neutrino masses, usually referred to as normal and inverted orderings. In the former, $\Delta m_{\rm sol}^2$ is the difference between the squared masses of the two lightest mass states, while in the latter it corresponds to the difference between the two heaviest states.

The \emph{Deep Underground Neutrino Experiment} (DUNE) \cite{Acciarri:2016crz, Acciarri:2015uup}, a new-generation long-baseline neutrino oscillation experiment, will perform precision measurements of the neutrino mixing parameters that govern $\nu_\mu \to \nu_e$ and $\overline{\nu}_\mu \to \overline{\nu}_e$ oscillations to fill in the missing pieces, namely:
\begin{itemize}
\item Measurement of the CP-violation phase, where a value differing from 0 or $\pi$ would represent the discovery of CP-violation in the leptonic sector, providing a possible explanation for the matter-antimatter asymmetry in the universe.
\item Determination of the neutrino mass ordering.
\item Precision tests of the three-flavour neutrino oscillation paradigm, including the measurement of the mixing angle $\theta_{23}$ and the determination of the octant in which this mixing angle lies.
\end{itemize}
The primary scientific program of DUNE includes as well proton-decay searches and neutrino astrophysics, topics discussed elsewhere in these proceedings \cite{Gil-Botella:2017}.

%%%%%%%%%%%%%%%%%%%%%%%%%%%%%%%%%%%%%%%%%%%%%%%%%%%%%%%%%%%%

\section{The DUNE detectors and the LBNF beam}
DUNE will consist of two detectors \cite{Acciarri:2016ooe} exposed to a new wide-band muon neutrino beam generated at the so-called \emph{Long-Baseline Neutrino Facility} (LBNF) beamline \cite{Strait:2016mof} that will be built at the Fermi National Accelerator Laboratory (Batavia, Illinois, USA). Neutrinos will be produced when protons extracted from Fermilab's Main Injector interact with a solid target to produce mesons which will be subsequently focused by magnetic horns into a 194~m long decay pipe where they decay into muons and neutrinos. The neutrino beam is aimed 4850~ft ($\sim1500$~m) underground at the Sanford Underground Research Facility (SURF) in South Dakota, about 1300~km away, where the so-called far detector (FD) will be located. The LBNF beamline is designed for initial operation at a proton-beam power of 1.2~MW, with the capability to support an upgrade to 2.4 MW. The parameters of the facility were determined taking into account the physics goals, spatial and radiological constraints, and the experience gained by operating the NuMI beam at Fermilab. 

%%%%%%%%%%
\begin{figure}[tb]
\centering
\includegraphics[width=\textwidth]{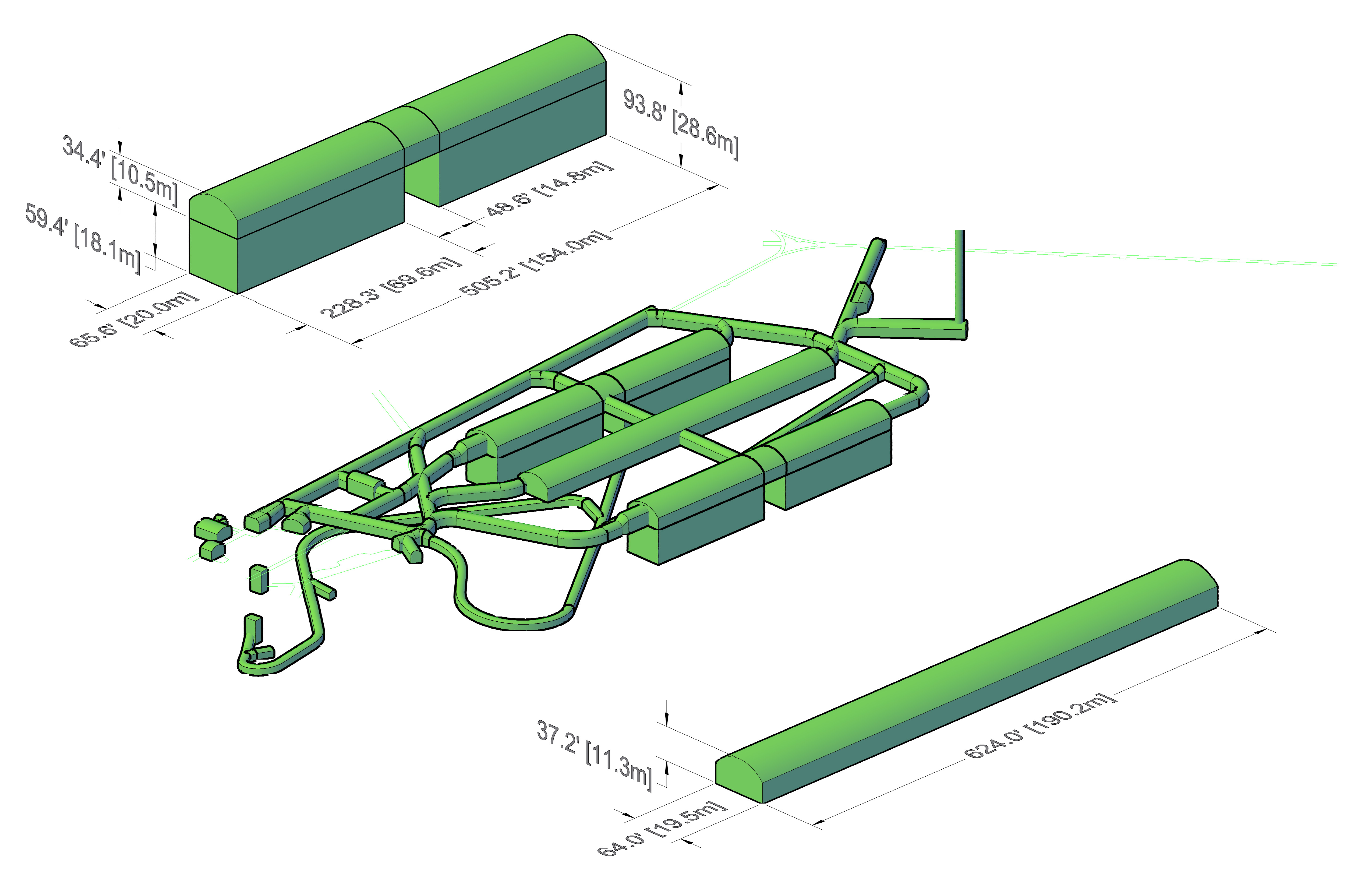} 
\caption{Approximated dimensions of the caverns that will be excavated at SURF to host the DUNE FD.} \label{fig:FD} 
\end{figure}
%%%%%%%%%%

The FD will be composed of four gigantic 10-kiloton liquid argon time projection chambers (LArTPC), with the dimensions shown in Figure~\ref{fig:FD}. The working principle of these detectors is as follows: charged-particles propagating through the LAr produce ionization and scintillation. The prompt scintillation light, registered by photo-detectors, provides the absolute time of the event. The ionization electrons are drifted with a constant electric field towards a segmented readout plane that records their amplitude and transverse positions. The longitudinal coordinate is obtained from the arrival time of the signals. Therefore, the LArTPC provides a three-dimensional measurement of the trajectory and energy-deposition pattern of the detected particles. The technology provides excellent tracking and calorimetry performance, and thus it is ideal for massive neutrino detectors such as the DUNE far detector, which require high signal efficiency and effective background discrimination, an excellent capability to identify and precisely measure neutrino events over a wide range of energies, and an excellent reconstruction of the kinematical properties with a high resolution. The full imaging of events will allow study of neutrino interactions and other rare events with an unprecedented resolution. The huge mass will grant the collection of a vast number of events, with sufficient statistics for precision studies.

The DUNE Collaboration is developing two different designs for the FD. The reference design adopts a single-phase readout, where the readout planes are immersed in the LAr volume. The alternative design is based on a dual-phase configuration, in which the ionization charges are extracted, amplified and detected in a layer of gaseous argon above the liquid volume. In the reference design, a 10-ktonne TPC module has an active volume 12~m high, 14.5~m wide and 58~m long. The TPC is instrumented with anode plane assemblies (APAs), which are 6.3~m high and 2.3~m wide, and cathode plane assemblies (CPAs) 3~m high by 2.3~wide. They are arranged in stacks forming walls (three CPAs interleaved by two APAs) and providing drift modules separated by 3.6~m each along the beam direction. The CPAs are held at $-180$~kV, such that ionization electrons drift a maximum distance of 3.6 m in an electric field of 500~$\mathrm{V/cm}$. The ultimate validation of the engineering solutions for both designs of the FD is foreseen in the context of the neutrino activities at CERN, where full-scale engineering prototypes are being assembled and commissioned \cite{Abi:2017aow}.  

To meet the ultimate systematic precision needed to fulfil the DUNE physics goals, another detector will be installed about 600 meters away from the start of the neutrino beamline. This so-called near detector (ND) must precisely characterize the neutrino beam energy and composition and measure to unprecedented accuracy the cross sections and particle yields of the various neutrino scattering processes. As the ND will be exposed to an intense flux of neutrinos, it will collect an extraordinarily large sample of neutrino interactions, allowing for an extended science program, including searches for heavy sterile neutrinos or non-standard interactions. The DUNE ND is currently under design, with several detector technologies being considered.

%%%%%%%%%%%%%%%%%%%%%%%%%%%%%%%%%%%%%%%%%%%%%%%%%%%%%%%%%%%%

\section{Sensitivity of DUNE}
%%%
DUNE plans to carry out a detailed study of neutrino mixing, resolve the neutrino mass ordering and search for CP violation in the lepton sector by studying the oscillation spectra of high-intensity $\nu_{\mu}$ and $\overline\nu_{\mu}$ beams measured over a long baseline. The electron neutrino appearance probability, P($\nu_{\mu} \to \nu_e$), is sensitive to both the value of the $\delta$ phase and the neutrino mass ordering. It shows a neutrino-antineutrino asymmetry introduced by both CP violation (for $\delta\neq0$ or $\pi$) and matter effects (the origin of the latter effect is simply the presence of electrons and absence of positrons in the Earth). In the few-GeV energy range, the asymmetry from matter effects increases with baseline as the neutrinos pass through more matter during their propagation, increasing as well the sensitivity to the neutrino mass ordering. For baselines longer than $1000$~km, the degeneracy between the asymmetries from matter and CP-violation effects can be resolved. The experiment must be capable of not only measuring the rate of electron neutrino appearance, but also of mapping out the oscillation energy spectrum down to energies of at least 500~MeV. In this way, DUNE will be able to unambiguously determine the neutrino mass ordering and measure the value of $\delta$. 

The experimental sensitivities presented here are estimated using GLoBES \cite{Huber:2004ka, Huber:2007ji}. GLoBES takes neutrino beam fluxes, cross sections and a detector-response parameterization as inputs. The cross-section inputs were generated using GENIE \cite{Andreopoulos:2009rq, Andreopoulos:2015wxa}. The neutrino oscillation parameters and their uncertainty were taken from the Nu-Fit \cite{Esteban:2016qun} global fit to neutrino data. Sensitivities to the neutrino mass ordering and the degree of CP violation are obtained by simultaneously fitting four different oscillated spectra: $\nu_\mu \to \nu_\mu$, $\nu_\mu \to \nu_e$, $\overline{\nu}_\mu \to \overline{\nu}_\mu$ and $\overline{\nu}_\mu \to \overline{\nu}_e$. 

\begin{figure}[p]
\centering
\includegraphics[clip, trim=1cm 0cm 1cm .75cm, width=.48\textwidth]{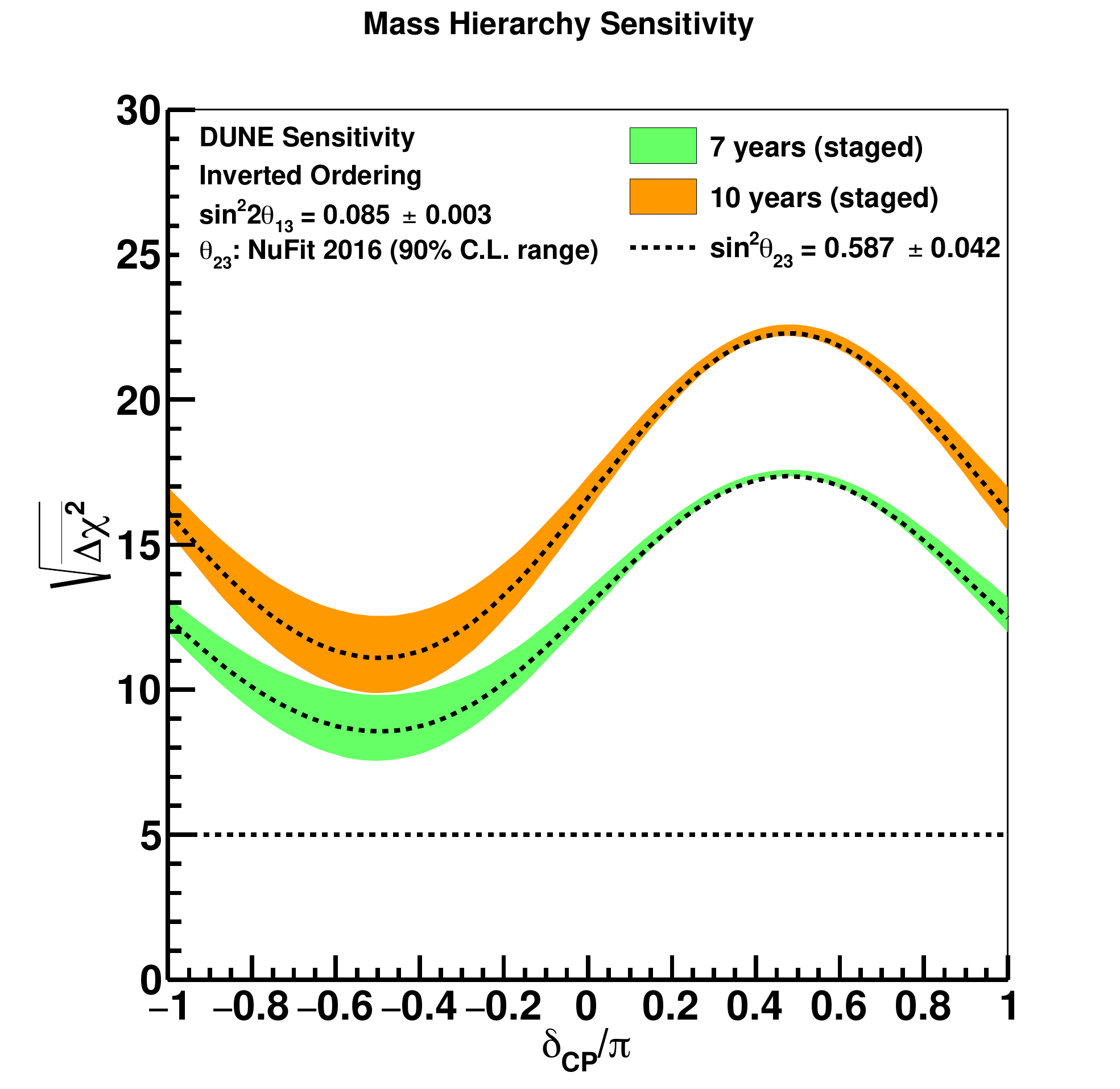} 
\includegraphics[clip, trim=1cm 0cm 1cm .75cm, width=.48\textwidth]{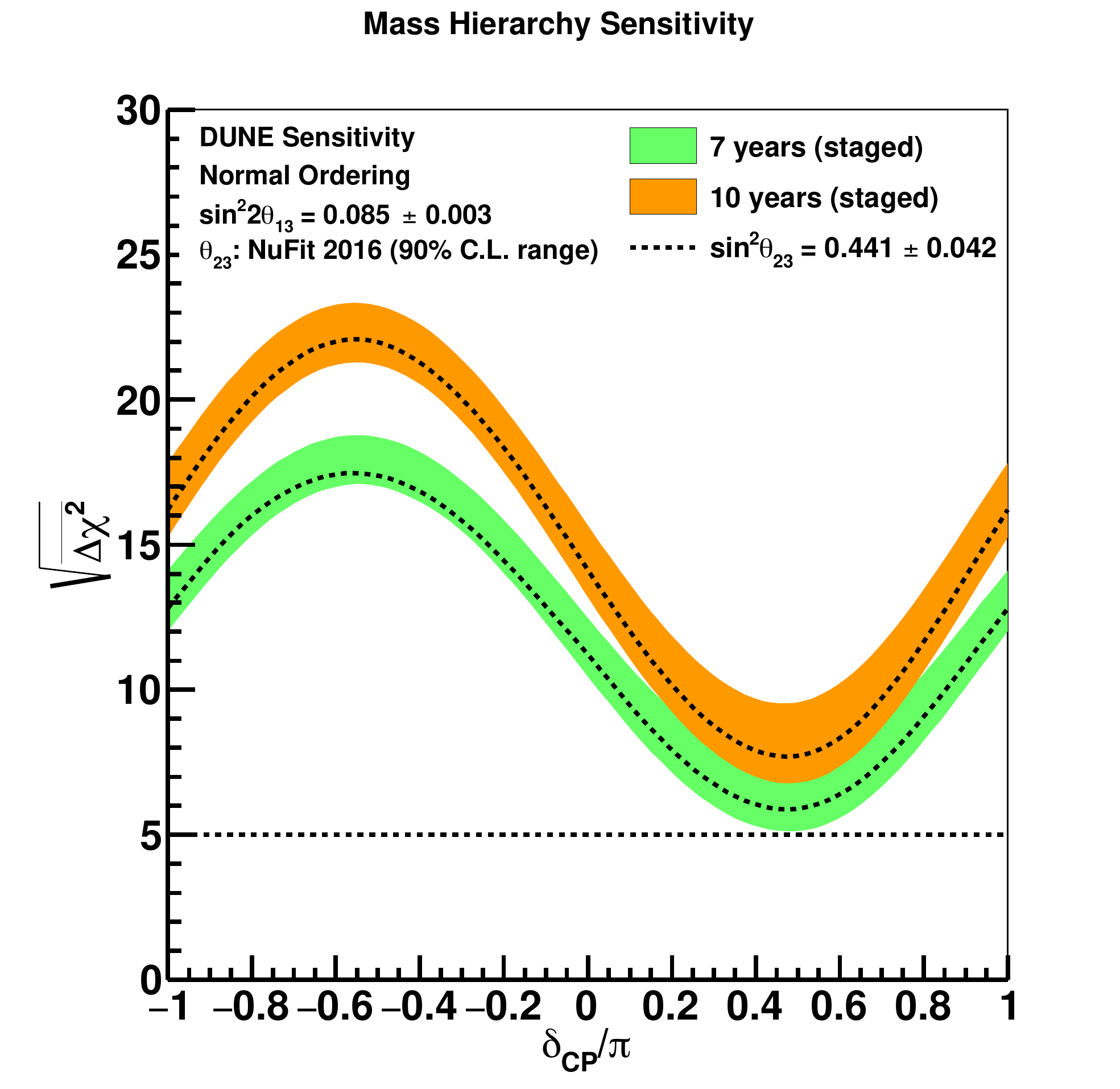} 
\caption{Sensitivity of DUNE to the neutrino mass ordering as a function of the true value of $\delta$ for exposures of seven (green) or ten (orange) years (in agreement with the DUNE staging scenario) for equal running in neutrino and antineutrino mode. In the left (right) panel, a true inverted (normal) ordering is assumed. The dashed line is the sensitivity for the NuFit central value of $\theta_{23}$ and the width of the band represents the range of sensitivities for the 90\% C.L. range in $\theta_{23}$ values.} \label{fig:MH} 
\end{figure}

\begin{figure}[p]
\centering
\includegraphics[clip, trim=1cm 0cm 1cm .75cm, width=.48\textwidth]{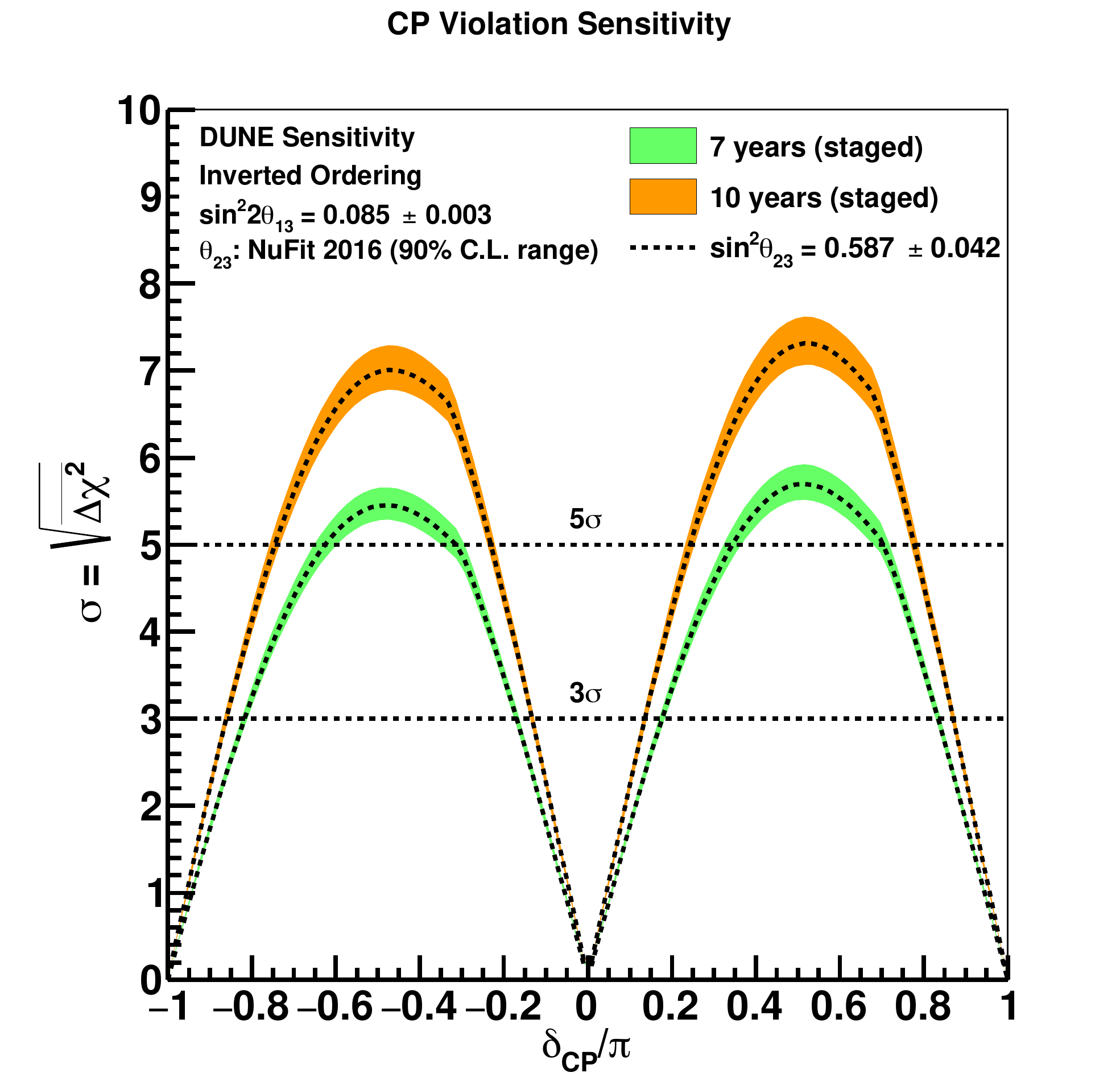} 
\includegraphics[clip, trim=1cm 0cm 1cm .75cm, width=.48\textwidth]{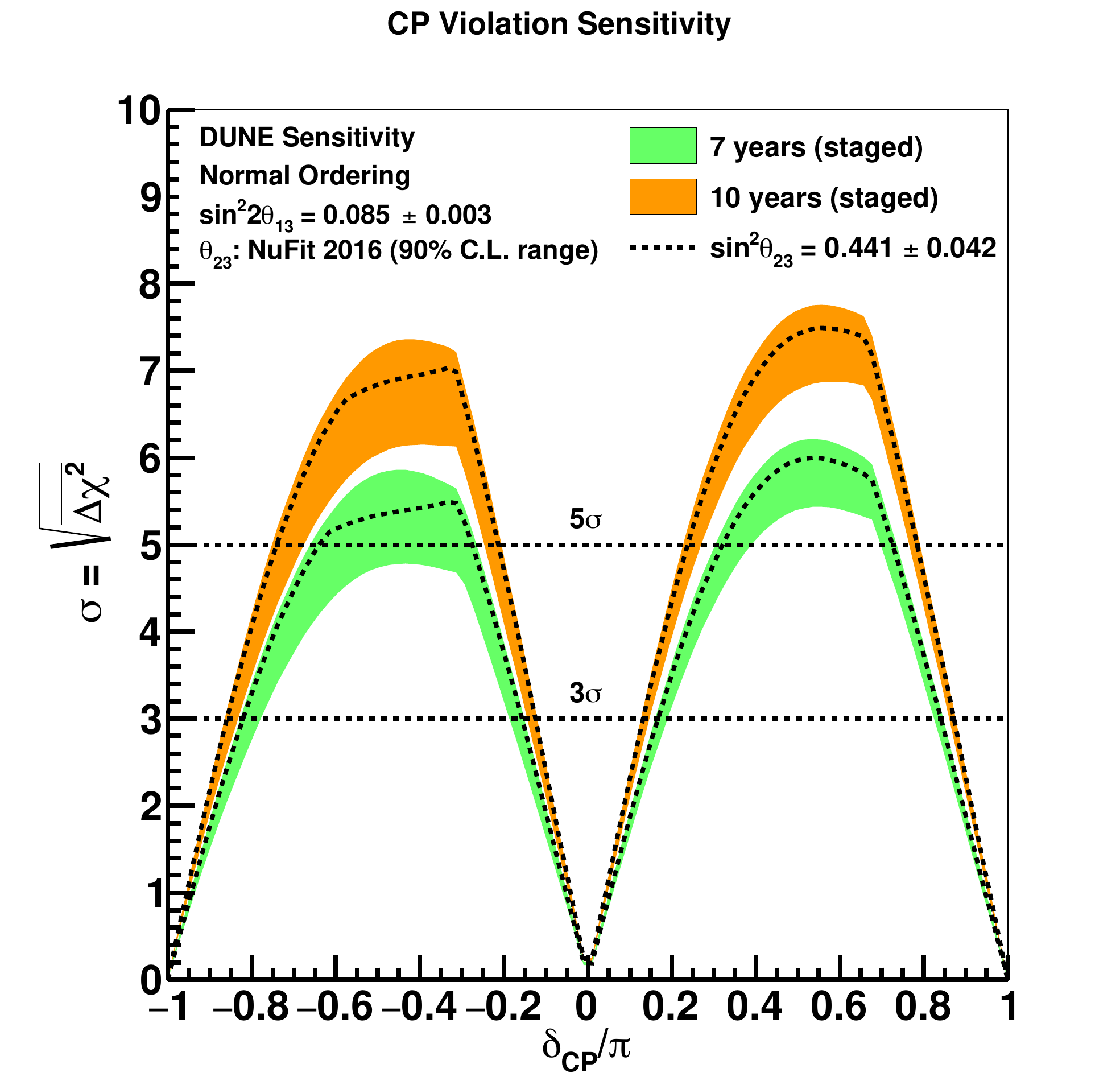} 
\caption{Sensitivity of DUNE to CP violation as a function of the true value of $\delta$ for exposures of seven (green) or ten (orange) years (in agreement with the DUNE staging scenario) for equal running in neutrino and antineutrino modes. In the left (right) panel, true inverted (normal) ordering is assumed. The dashed line is the sensitivity for the NuFit central value of $\theta_{23}$ and the width of the band represents the range of sensitivities for the 90\% C.L. range in $\theta_{23}$ values.} \label{fig:CP} 
\end{figure}

Figure~\ref{fig:MH} shows the significance with which the MH can be determined as a function of the value of $\delta_{CP}$, for an exposure which corresponds to seven years of data (3.5 years in neutrino mode plus 3.5 years in antineutrino mode) with a 40-ktonne detector and a 1.07-MW (80 GeV) beam. For this exposure, the MH is determined with a minimum significance of $\sqrt{\overline{\Delta\chi^2}}$= 5 for nearly 100\% of $\delta_{CP}$ values for the reference beam design. Figure \ref{fig:CP} shows the significance with which the CP violation ($\delta_{CP} \neq $ 0 or $\pi$) can be determined as a function of the value of $\delta_{CP}$ for an exposure of 300 kt $\times$ MW $\times$ year, which corresponds to seven years of data (3.5 years in neutrino mode plus 3.5 years in antineutrino mode) with a FD of 40 kitonnes and a 1.07-MW beam.

%%%%%%%%%%
%%%%%%%%%%

%%%%%%%%%%
%%%%%%%%%%

%%%%%%%%%%%%%%%%%%%%%%%%%%%%%%%%%%%%%%%%%%%%%%%%%%%%%%%%%%%%

\bibliographystyle{JHEP} 
\bibliography{references}

\end{document}